# INTERNAL STRUCTURE OF A GAUSS-BONNET BLACK HOLE

S.O. Alexeyev

*Department of Theoretical Physics, Physics Faculty, Moscow State University, Moscow 119899, Russia*

Black holes are studied in the frames of superstring theory using a non-trivial numerical integration method. A low energy string action containing graviton, dilaton, Gauss-Bonnet and Maxwell contributions is considered. Four-dimensional black hole solutions are studied inside and outside the event horizon. The internal part of the solutions is shown to have a non-trivial topology.

## 1. Introduction

In the recent years there was a heated discussion about the nature of dark matter. Among possible candidates there are black holes. Studying their properties in the frames of superstring theory, one can hope to clarify some aspects of the dark matter nature. That is one of the reasons for a great interest in investigations of the low energy string action in four dimensions. Some researches [1, 2, 3, 4, 5, 6] found that the well-known solutions (such as the Schwarzschild one or Gibbons-Maeda-Garfincle-Horowitz-Strominger (GM-GHS) one) should be modified by higher order curvature corrections.

In our previous work [5] the internal structure of black hole solutions for the Lagrange density $\mathcal{L} = m_{\rm Pl}^2(-R + 2\partial_\mu\phi\partial^\mu\phi) + \lambda\, {\rm e}^{-2\phi} S_{GB}$ were studied. It is of interest to find the influence of the Maxwell term on black hole solutions of 4D low energy string gravity with the second-order curvature corrections. Some researchers [2, 6] consider the bosonic part of the gravitational action consisting of dilaton, graviton, Maxwell and Gauss-Bonnet (GB) terms (for simplicity, the antisymmetric tensor terms are ignored) taken in the following form:

$$S = \frac{1}{16\pi}\int d^4x\sqrt{-g}\bigg[m_{\rm Pl}^2(-R + 2\partial_\mu\phi\partial^\mu\phi) - {\rm e}^{-2\phi}F_{\mu\nu}F^{\mu\nu} + \lambda\, {\rm e}^{-2\phi} S_{GB}\bigg], \qquad (1)$$

[1]e-mail: alexeyev@grg2.phys.msu.su

where $R$ is the scalar curvature, $\phi$ is the dilaton field, $m_{\rm Pl}$ is the Planck mass; $F_{\mu\nu}F^{\mu\nu}$ is the Maxwell field and $\lambda$ is the string coupling parameter. The last term describes the GB contribution ($S_{GB} = R_{ijkl}R^{ijkl} - 4R_{ij}R^{ij} + R^2$) to the action (1). Such configurations were partly studied [2] by the perturbative analysis $O(\lambda)$ outside the event horizon ($r_h$) when $r_h \gg m_{\rm Pl}$. The authors showed that the black hole solution are real and provide non-trivial dilatonic hair. The solutions beyond the event horizon are very important from the viewpoint of quantum gravity because it is generally believed [7] that in the regions of space-time with sufficiently small curvature a *classical* solution gives the main contribution to the global structure of the space-time. Quantum corrections may drastically modify the properties of space when the curvature is large enough. A study of complete black hole solutions for the action (1) is the aim of this work.

## 2. Field equations

The aim is to find static, asymptotically flat, spherically symmetric black-hole-like solutions. In this case the most convenient choice of the metric is

$$ds^2 = \Delta dt^2 - \frac{\sigma^2}{\Delta}dr^2 - f^2(d\theta^2 + \sin^2\theta d\varphi^2), \qquad (2)$$

where the functions $\Delta$, $\sigma$ and $f$ depend only on the radial coordinate $r$. Substituting the expressions of $R$ and $S_{GB}$ into the action (1) and integrating this modified action by parts and over the angle variables, one can rewrite it in a somewhat more convenient form (for the present analysis the boundary term is not relevant and is ignored)

$$S = \frac{1}{2}\int dt dr \left[ m_{\rm Pl}^2 \left( \frac{\Delta' f' f}{\sigma} + \frac{\Delta (f')^2}{\sigma} \right. \right.$$
$$\left. + \sigma - \frac{\Delta f^2 (\phi')^2}{\sigma} \right) - e^{-2\phi} q^2 \frac{\sigma}{f^2}$$
$$\left. + 4 e^{-2\phi} \lambda \phi' \left( \frac{\Delta \Delta' (f')^2}{\sigma^3} - \frac{\Delta'}{\sigma} \right) \right]. \quad (3)$$

We will consider a black hole with a purely magnetic charge, so that the Maxwell tensor $F_{\mu\nu}$ can be written in the form $F = q \sin\theta \, d\theta \wedge d\varphi$. The corresponding field equations in the GHS gauge $[\sigma(r) = 1]$ are

$$m_{\rm Pl}^2 (f'' f + f^2 (\phi')^2)$$
$$+ 4 e^{-2\phi} \lambda [\phi'' - 2(\phi')^2][\Delta (f')^2 - 1]$$
$$+ 4 e^{-2\phi} \lambda \phi' 2 \Delta f' f'' = 0,$$

$$m_{\rm Pl}^2 (1 + \Delta f^2 (\phi')^2 - \Delta' f' f - \Delta (f')^2)$$
$$+ 4 e^{-2\phi} \lambda \phi' \Delta' (1 - 3\Delta (f')^2) - e^{-2\phi} q^2 f^{-2} = 0,$$

$$m_{\rm Pl}^2 [\Delta'' f + 2\Delta' f' + 2\Delta f'' + 2\Delta f (\phi')^2]$$
$$+ 4 e^{-2\phi} \lambda [\phi'' - 2(\phi')^2] 2 \Delta \Delta' f'$$
$$+ 4 e^{-2\phi} \lambda \phi' 2((\Delta')^2 f' + \Delta \Delta'' f' + \Delta \Delta' f'')$$
$$- 2 e^{-2\phi} q^2 f^{-3} = 0,$$

$$- 2 m_{\rm Pl}^2 [\Delta' f^2 \phi' + 2\Delta f f' \phi' + \Delta f^2 \phi'']$$
$$+ 4 e^{-2\phi} \lambda ((\Delta')^2 (f')^2 + \Delta \Delta'' (f')^2$$
$$+ 2\Delta \Delta' f' f'' - \Delta''] - 2 e^{-2\phi} q^2 f^{-2} = 0. \quad (4)$$

It is necessary to note that the GM-GHS solution

$$ds^2 = \left[ 1 - \frac{2M}{r} \right] dt^2 - \left[ 1 - \frac{2M}{r} \right]^{-1} dr^2$$
$$- r \left[ r - \frac{q^2 \exp(2\phi_0)}{M} \right] d\Omega,$$
$$\exp(-2\phi) = \exp(-2\phi_0) - \frac{q^2}{Mr}, \quad (5)$$

is the basic solution for the $U(1)$ purely magnetic case (when $\lambda = 0$). If $F = 0$ in (1), the basic solution is the well-known Schwarzschild one with a constant dilaton field (according to the "no-hair" theorem). Moreover, the solution of Eqs. (4) at infinity must have the GM-GHS form.

## 3. Numerical Results

For integrating inside the event horizon a method based on integration over an additional parameter was used as described in our previous paper [5]. The main result of that work is the following. An asymptotically flat black hole solution for the action (1) without the Maxwell term exists from infinity down to the end point $r = r_s$ (see Fig. 1 in Ref. [5]) inside the regular event horizon ($r_h$). When $r_h$ is large enough (the contribution of the second order curvature corrections is small) the position of the end point of the solution is $r_s \ll r_h$. As $r_h$ decreases (the GB term contribution increases), the distance between $r_s$ and $r_h$ becomes smaller and smaller. The curvature invariant also diverges near the position $r = r_s$. An additional (nonphysical) branch of the solution begins at the point $r_s$ and exists up to a point $r_x$ which may be called a singular horizon. There is no other solution in the neighborhood of $r_s$.

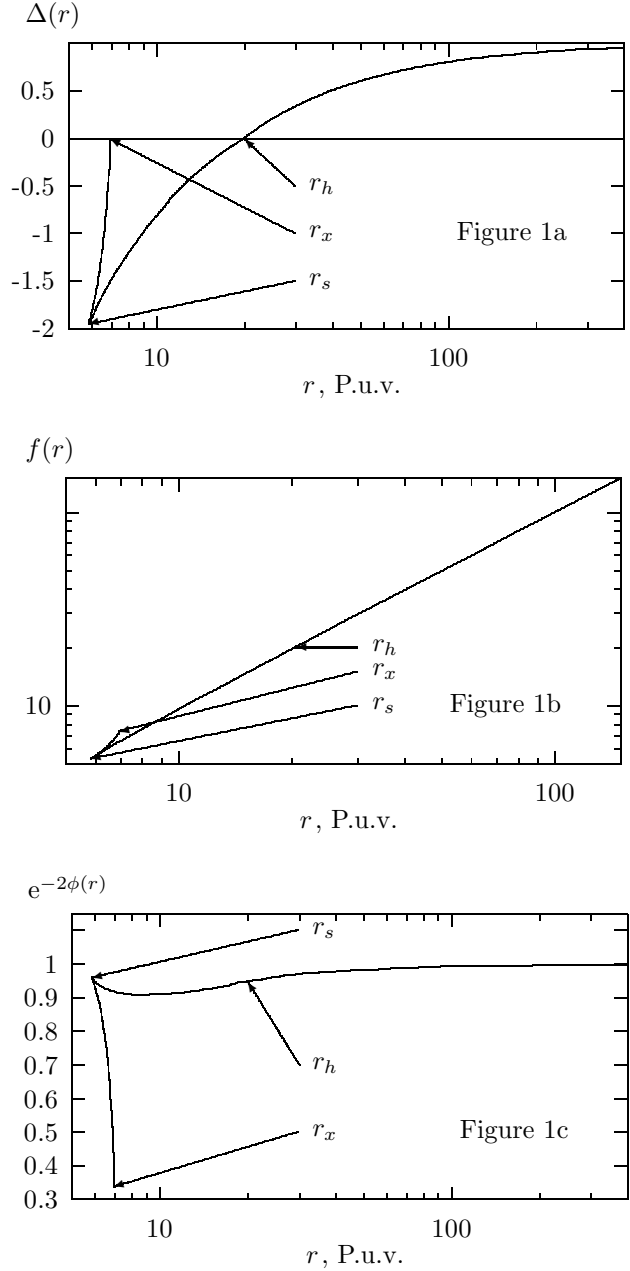

Figure 1: The dependence of the metric functions $\Delta$ (a), $f$ (b) and the dilaton function $e^{-2\phi}$ (c) on the radial coordinate $r$ when the event horizon radius $r_h$ is equal to 20.0 Planck unit values (P.u.v.) and the magnetic charge is $q < q_{cr}$.

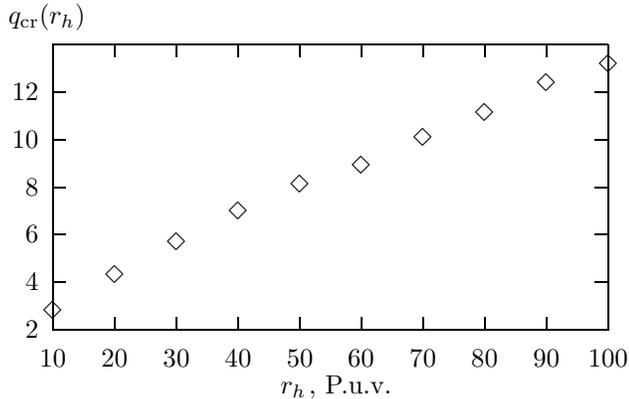

Figure 2: Critical magnetic charge $q_{rmcr}$ vs. regular event horizon radius $r_h$ for $\lambda = 1$.

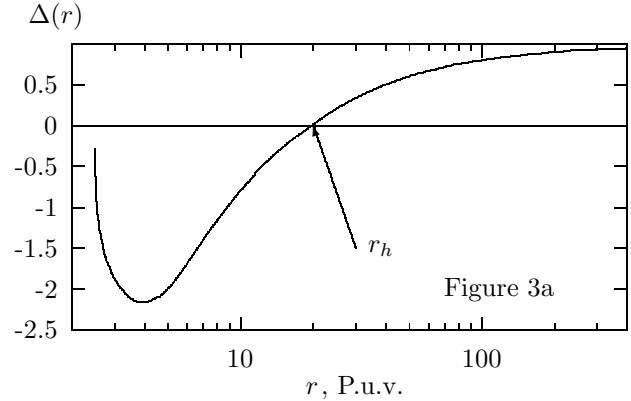

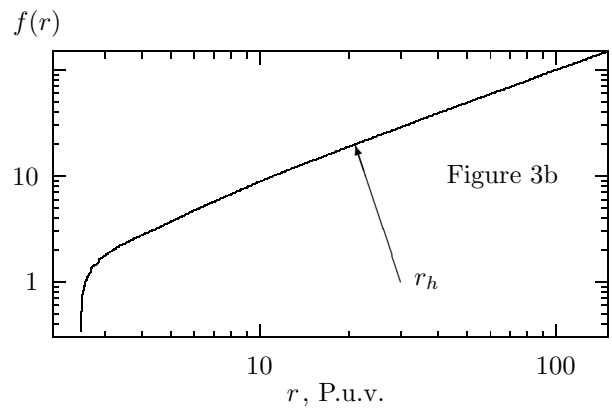

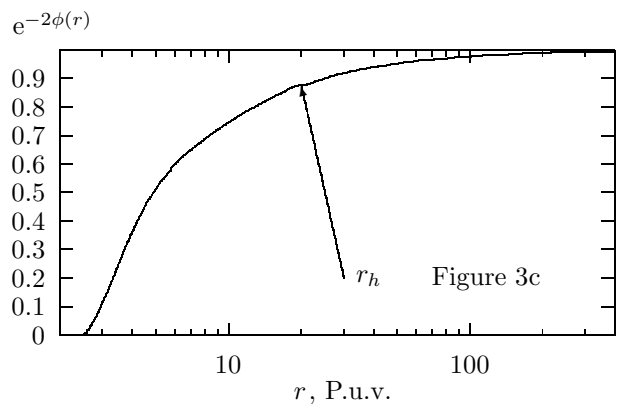

Figure 3: The dependence of the metric functions $\Delta$ (a), $f$ (b) and dilaton function $\exp(-2\phi)$ (c) on the radial coordinate $r$ when the event horizon radius $r_h$ is equal to 20.0 Planck unit values (P.u.v.) and the magnetic charge is $q > q_{rmcr}$.

When one includes the Maxwell term in the action (1), the resulting picture is as follows. Black hole solutions of Eqs. (4) exist only in the range of the magnetic charge values $0 \leq q \leq m\sqrt{2}$, as in the GM-GHS case. The solution behaviour outside the regular event horizon $r_h$ looks like the GM-GHS one, which coincides with the results of Mignemi [2] and Maeda [6]. The solutions behaviour inside the regular event horizon $r_h$ depends on the magnetic charge $q$. When $q$ is quite small, the solutions have a form analogous to the purely Einstein-dilaton-Gauss-Bonnet (EDGB) case (see Fig. 1) because the contribution of the second-order curvature term is "stronger" than the Maxwell one. The curvature invariant $R_{ijkl}R^{ijkl}$ and the components $T_0^0$ and $T_2^2$ of the stress-energy tensor diverge near the position $r_s$. Hence one can conclude that the point $r_s$ represents a "pure scalar singularity" (see e.g. the classifications of space-time singularities in [8, 9]). With increasing $q$ the contribution of the Maxwell term becomes greater and greater and at some $q = q_{cr}$ the behaviour of the solution changes and the turning point $r_s$ disappears. The dependence of the critical magnetic charge value on the event horizon radius $r_h$ is depicted in Fig. 2. It is necessary to note that, when $q < q_{cr}$, the quantity $q$ is rather small and in all the particular points (namely, $r_h$, $r_s$, $r_x$) one can use asymptotic expansions obtained for the pure EDGB case [5]. When $q > q_{cr}$, the main asymptotically flat branch ($\infty \ldots r_h \ldots r_s$) and the additional nonphysical branch ($r_s \ldots$) merge to form a single asymptotically flat black hole branch that exists from infinity down to the point where the metric function $f$ vanishes. The function $f$ and the dilatonic function $e^{-2\phi}$ have the same behaviour as in the GM-GHS case. The metric function $\Delta$ exhibits a local minimum not far from the position where $f$ vanishes — this feature is absent in the GM-GHS solution (see Fig. 3).

## 4. Discussion and conclusions

In this work we have obtained black hole solutions with nontrivial dilatonic "hair" for low-energy effective string action with second-order curvature corrections and a Maxwell magnetic field.

The solutions are characterized by the ADM mass $M$, the dilaton charge $D$ and the asymptotic dilaton value $\phi_\infty$. They are stable under fluctuations of initial conditions. Since these solutions have a nonperturbative nature, they are not restricted by any perturbative parameter values.

One interesting result of this work is the coincidence of the solutions of EDGB and EDGB $+ \, \mathrm{e}^{-2\phi} F^2$ systems in the case $q < q_{\mathrm{cr}}$ and the appearance of the $r_s$ singularity inside the black hole after adding the GB term to the action. This singularity has the topology $S^2 \times R^1$, i.e. it is an infinite (in $t$ direction) "tube" of radius $r_s$. A similar "tube" in the Schwarzschild metric with an additional condition of $R^{abcd}R_{abcd}$ finiteness was discussed by V. Frolov et al. [7]. There are two solutions on this "tube". The asymptotically flat solution, which is the main one, starts from $r_s$ and continues to infinity. Outside the regular event horizon the solution looks like the GM-GHS one, which agrees with the results of Mignemi [2], Maeda [6] and Kanti [3]. The additional nonphysical solution branch provides the existence of a "singular" inner horizon with $R_{ijkl}R^{ijkl} \to \infty$. Some solutions exist inside the "tube" $r_s$, but they are unstable under initial condition fluctuations, and we cannot distinguish, which branch, main or additional, they correspond to. In the case $q > q_{\mathrm{cr}}$ the solution looks like the GM-GHS one outside and inside the regular event horizon $r_h$, with an additional local maximum of the metric function $\Delta$.

### Acknowledgement

The author would like to thank Professor D.V. Gal'tsov for useful discussions of the subject of this work.